\documentclass[twocolumn]{jpsj2}

\title{Nonlocal Manipulation of Dimer Motion at Ge(001) Clean Surface via Hot Carriers in the Surface States}

\author{Yasumasa \textsc{TAKAGI}\thanks{Present address: RIKEN/SPring-8, Mikazuki-cho, Hyogo 679-5148, Japan, E-mail address: y-takagi@spring8.or.jp}, Yoshihide \textsc{YOSHIMOTO}, Kan \textsc{NAKATSUJI} and Fumio \textsc{KOMORI}\thanks{E-mail address: komori@issp.u-tokyo.ac.jp}}

\inst{Institute for Solid State Physics, University of Tokyo, 5-1-5 Kashiwanoha, Kashiwa-shi, Chiba 277-8581, Japan}

\abst{Nonlocal one-dimensional motions of a topological defect are induced by electron tunneling through the dangling-bond states on the clean Ge(001) surface using scanning tunneling microscopy below 80 K. The direction of the motion depends both on the energy of the carriers in the surface state and on the distance between the defect and the tunneling point. The results are interpreted using an electronic excitation model by hot carriers injected to the surface states. The critical distance of the motion is anisotropic and consistent with the band structure of the surface states.}

\kword{Surface Hot Carrier, Atom Manipulation, Germanium, Scanning Tunneling Microscopy}

\begin{document}
\maketitle

Local electron tunneling often induces motions of atoms and molecules at surfaces, and their dynamics has been studied using scanning tunneling microscopy (STM). In most of the cases, the electron is directly injected to or extracted from their localized electronic states just under the tip, and the electronic energy is transferred to the kinetic energy of atoms or molecules by inelastic tunneling processes \cite{Eigler,Lyding,Stipe}. On semiconductor surfaces, the lifetime of the surface electron is longer than on metal surfaces, and a nonlocal excitation through a surface state has been expected. Previously, nonlocal motions of the halogen adsorbates by the tunneling were reported on Si(111) \cite{Maeda} and Si(001) \cite{Weaver} surfaces. However, the propagation of the injected hot carrier has not been studied in detail. It is important to understand the mechanism of such nonlocal manipulations both for fundamental physics on electron and lattice dynamics and for their application to nanostructure fabrications.

In the present letter, we show motions of a topological defect on the clean Ge(001)-$c$(4$\times$2) and -$p$(2$\times$2) surfaces at 80 K as functions of the distance and the direction between the tunneling point and the defect (cf. Fig. 1). The defect can be easily created and annihilated by tunneling current \cite{Takagi4}, and thus is suitable for the quantitative study of the nonlocal manipulation.  Its motion is anisotropically induced by electron injection to the surface while the excitation is isotropic in the case of hole injection. The critical distance of the motion can be more than 80 nm by the electron injection. Moreover the direction of the motion depends on the sample bias voltage and the local superstructure.
The observed anisotropy of the defect motion can be understood by considering the dispersion relations of the surface electronic bands.

\begin{figure}[htbp]
\begin{center}
\includegraphics{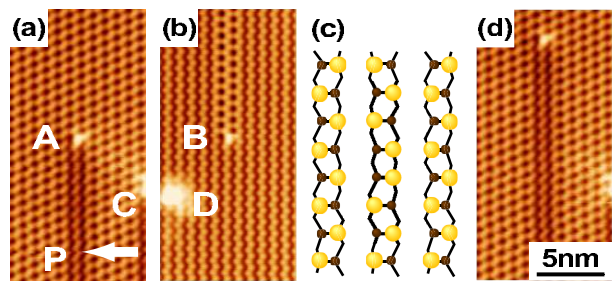}
\end{center}
\caption{(color online)(a,b) STM images including topological defects, A, B, which were created on a $c$(4$\times$2) surface (a), or on a $p$(2$\times$2) surface (b). Protrusions at C in (a) and D in (b) are impurities as fixed markers. (c) A schematic model of the defect between the two superstructures. (d) STM image after the bias pulse from $-$ 0.4 V to 0.8 V for 1 sec at P in (a). The STM tip position is fixed during the pulse.}
\label{fig:1 }
\end{figure}

The Ge(001) clean surface consists of the buckled dimers, and its buckling orientation is alternate in the direction perpendicular to the dimer axis. We call the row of the dimers in this direction a ``dimer row". There are two stable arrangements of the dimers; one is the $c$(4$\times$2) structure, and the other is $p$(2$\times$2). Figures 1(a,b) show STM images at 80 K including a narrow $p$(2$\times$2) area on $c$(4$\times$2) surface (a), and a $c$(4$\times$2) area on $p$(2$\times$2) surface (b). At the boundary between the two structures in the dimer row, there are two adjacent dimers with the same buckling orientation as a topological defect. Its model is schematically illustrated in Fig. 1(c). It is imaged higher than the other Ge atoms on the surface when the sample bias voltage $V_{b}$ is negative. Here, we call it a ``kink". For simplicity, we also define the kink shown in Fig.1(a) as A-type and that in Fig. 1(b) as B-type. 

The STM images were observed at 80 K in an ultrahigh vacuum system \cite{Naitoh}. The $c$(4$\times$2) Ge(001) clean surface was obtained by several repetitions of Ar ion sputtering and annealing. We made a large area of the $p$(2$\times$2) structures by scanning the surface with $V_{b}$ = 1.2 V and the tunneling current $I_{t}$ = 1 nA \cite{Takagi4,Takagi1}, typically. We can convert the $p$(2$\times$2) surface to $c$(4$\times$2) again by scanning the surface with $V_{b}$ = $-$ 1.2 V.

The kink is created and moved on the surface by injecting electrons or holes at its neighboring point using STM below 80 K. For example, the A-type kink was made by changing $V_{b}$ from $-$ 0.4 V to 0.8 V for 3 sec with fixing the tip position over a dimer in the $c$(4$\times$2) surface. Generally, a pulse operation creates two kinks, and there forms a one-dimensional area of $p$(2$\times$2) between them. In the present study, the one end of the area was fixed at a step edge, and we focused on the motion of the kink on the terrace. The surface can be clearly imaged without moving the kink using $V_{b} = -$ 0.4 V and $I_{t}$ = 1 nA.
\begin{figure}[htbp]
\begin{center}
\includegraphics{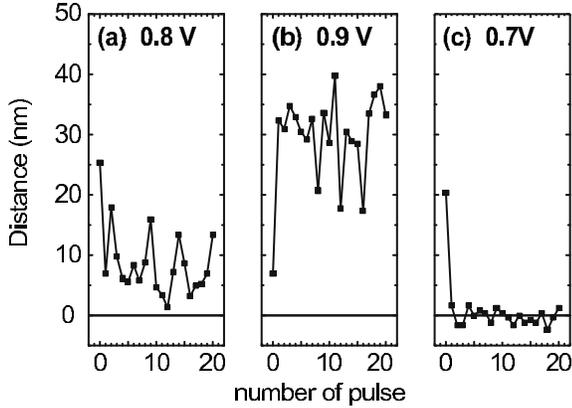}
\end{center}
\caption{The distance between the kink and the tunneling point in the dimer-row direction after each bias pulse from $-$ 0.4 V to 0.8 V (a), 0.9 V (b) and 0.7 V (c) for 1 sec was applied at the point on the dimer row including the kink. The current during each pulse was 25 nA (a), 33 nA (b) and 17 nA (c).}
\label{fig:2}
\end{figure}

A $V_{b}$ pulse at the position of P on the surface shown in Fig. 1(a) randomly moves the A-type kink in the dimer-row direction without any other changes in the surface structure as in Fig. 1(d). We observed STM images with $V_{b}$ = $-$ 0.4 V after each pulse to 0.8 V for 1 sec. Figure 2(a) summarizes the random motion of the kink position by the pulse along the dimer row. The center of the random motion is 10 nm apart from the position of the tunneling. The separation between the center and P increases with increasing $V_{b}$ during the pulse. It is 30 nm with the pulse to 0.9 V as in Fig. 2(b), and over 50 nm with the pulse to 1.0 V. On the other hand, when we use the pulses to 0.7 or 0.6 V, the kink first comes close to the tunneling point, and moves randomly around the point as in Fig. 2(c) for the 0.7 V pulse. The width of the random motion by the 0.6 V pulse is smaller than that by the 0.7 V pulse, and the positive pulse applied to $V_{b}$ of less than 0.5 V never induces the kink motion. 
In the case of the B-type kink, the distance between the tunneling point and the center of the random walk rapidly increases with increasing $V_{b}$ of the pulse above 0.7 V, and becomes more than 50 nm with the pulse to 0.75 V for 1 sec.

Random motion of the kink by a negative bias pulse was previously reported \cite{Takagi4}. The distance between the center of the kink motion and the tunneling point increases with decreasing $V_{b}$ of the pulse below $-$ 0.7 V. In the negative bias case, there is no bias range where the kink comes back to the tunneling point; changing $V_{b}$ of the pulse to above $-$ 0.6 V does not move the kink. The preferred direction of the kink motion, thus, largely depends on the bias voltage and the distance from the tunneling point.

\begin{figure}[htbp]
\begin{center}
\includegraphics{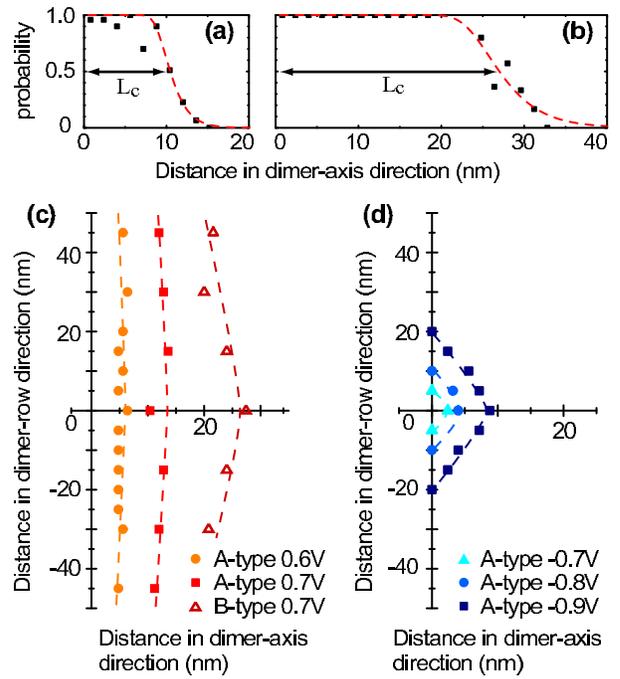}
\end{center}
\caption{(color online) (a,b) Probability of the kink motion per bias pulse as a function of the distance between the tunneling point for the A-type (a) and B-type (b) kinks. The pulse is from $-$ 0.4 V to 0.7 V for 1 sec. The dotted curves are obtained using eq. 1 with the parameters fitted to the data. (c) Two-dimensional plot of the tunneling point where the probability of the kink motion is 0.5 with positive pulses from $-$ 0.4 V to 0.6 and 0.7 V for the A-type kink, and to 0.7 V for the B-type kink. The kink is located at the origin. (d) The same plot with negative pulses from $-$ 0.4 V to $-$ 0.7, $-$ 0.8, and $-$ 0.9 V for the A-type kink.}
\label{fig:3}
\end{figure}

Motions of the kink were repeatedly observed as functions of the distance and direction from the tunneling point for the both types of the kink. We show in Figs. 3(a,b) the probability of the motion per pulse vs. the distance in the dimer-axis direction. The STM tip position is fixed with $V_{b}$ = $-$ 0.4 V and $I_{t}$ = 1 nA, and the pulse is to 0.7 V for 1 sec. In this condition, we injected 1.1 $\times$ $10^{11}$ electrons per pulse at the tunneling point. The probability of the kink motion rapidly decreases at a certain distance. We define here the critical distance of the motion, $L_c$, as the distance where the probability is 0.5. It depends on the type of the kink, the bias voltage, and the direction from the kink to the tunneling point. 

In Fig. 3(c), we two-dimensionally plot the points where the probability is 0.5 for the both types of the kink. The total number of injected electrons for the 0.6 V pulse was 8 $\times$ $10^{10}$. The critical distance is highly anisotropic and exceeds 80 nm in the dimer-row direction even for the pulse to 0.6 V. Thus, the ratio of $L_c$ in the dimer-row direction to that in the dimer-axis direction is more than ten. On the other hand, the ratio is two in the case of the hole injection to the surface as in Fig. 3(d). The total numbers of injected electrons were 2.5 $\times$ $10^{10}$ for the $-$ 0.7 V pulse, 4.4 $\times$ $10^{10}$ for $-$ 0.8 V, and  6.3 $\times$ $10^{10}$ for  $-$ 0.9 V.

The critical distance observed in the dimer-axis direction by electron injection depends on the type of the kink. To understand the origin of the difference, we prepared thin $c$(4$\times$2) areas in different ways between the B-type kink and the tunneling point on the $p$(2$\times$2) surface as in Figs. 4(a-d). We measured $L_c$ on these surfaces using the same pulse from $-$ 0.4 to 0.7 V. Here we always injected the current on the $p$(2$\times$2) surface to maintain the tunneling condition the same. We found little difference of $L_c$ between the surfaces shown in Figs. 4(b) and 4(c). These two surfaces have the same width of $c$(4$\times$2) area in total, but the arrangement of the two areas is different. This means that $L_c$ depends mainly on the total width $W$. The observed $L_c$ is almost a linear function of $W$ as in Fig. 4(e). 

\begin{figure}[htbp]
\begin{center}
\includegraphics{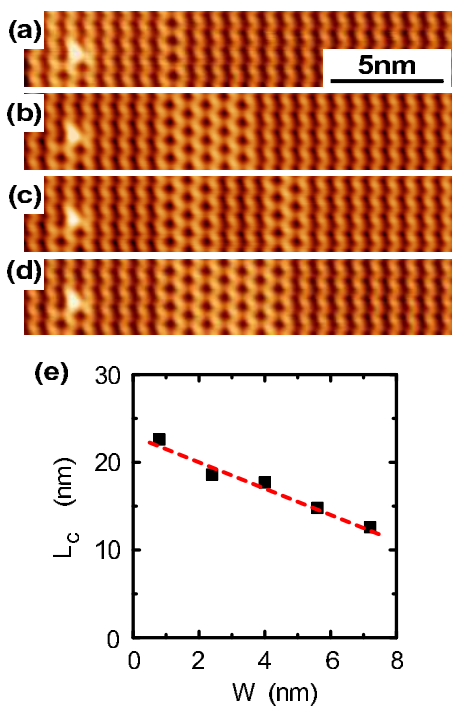}
\end{center}
\caption{(color online) (a-d) STM images showing the surfaces with different local structures used for the measurements of the B-type kink motions. The total width of the $c$(4$\times$2) area, $W$, is the same in (b) and (c). (e) The critical distance $L_c$ in the dimer-axis direction as a function of $W$. The slope of the dotted line is $-$ 1.5.
}
\label{fig:4 }
\end{figure}

For the discussion of the experimental results, we adopt the model that the kink motion is induced through an inelastic scattering process by the injected hot carriers in the substrate as in the case of the adsorbate motions \cite{Maeda,Weaver}. In the present case, the hot carrier maintains its energy enough to induce the kink motion during the propagation from the tunneling point to the kink in the surface state. The threshold energy of the kink motion is around 0.6 eV for both electron and hole injections. It is larger than the theoretically estimated barrier of the kink motion, 0.3 eV \cite{Kawai1}. The apparent discrepancy, however, can be removed if we consider the very small rate of the inelastic process through electron-phonon interaction below the observed threshold energy \cite{Kawai2}. 

The anisotropy of the critical distance depends on the electronic structures of the surface. The valence and conduction band structures of the Ge(001) surface have been studied by photoemission, inverse photoemission and tunneling spectroscopies \cite{Landermark1,Landermark2,Kipp,Nakatsuji}. The filled $\pi$ band is a nearly isotropic surface state, and is a resonance to the bulk valence band around $\overline{\Gamma}$. On the other hand, the empty $\pi^{*}$ band is highly anisotropic, and is isolated from the bulk states in the band gap. It strongly disperses in the dimer-row direction, and there is little dispersion in the dimer-axis direction. These experimental results are qualitatively consistent with the band calculations for the $c$(4$\times$2) surface \cite{Nakatsuji} and for the 2$\times$1 surface with the buckled dimers \cite{Rohlfing}. Consequently, the hole injected to the $\pi$ band two-dimensionally propagates in the surface. This explains the observed nearly isotropic $L_c$ in Fig. 3(d) while the anisotropic $L_c$ in Fig. 3(c) for the electron injection is attributed to the anisotropy of the hot electron propagation in the $\pi^{*}$ band. 

The relaxation time of the hot carriers injected to the surface bands is crucial in the present model. It looses the energy through electron-phonon and electron-electron interaction. The dynamics of the hot electron in the $\pi^{*}$ band
 has not been studied for the clean Ge(001) surface, but was measured by time-resolved two-photon photoemission spectroscopy for the Si(001) surface \cite{Weinelt}, which has a similar electronic structure to the Ge(001) surface around the Fermi energy. The hot electron in the $\pi^{*}$ band of the Si(001) surface inelastically scatters within the band via phonon emission to the band bottom at $\overline{\Gamma}$ in 1.5 ps. After that, it further relaxes from the band bottom to the other states including the bulk states. Furthermore, a time scale of 50 fs was reported for the intraband inelastic scattering time via phonon emission using a five-wave-mixing method \cite{Voelkmann}. The short scattering time was explained by a strong electron-phonon interaction at the surface. On the Ge(001) surface, a similar intraband relaxation process of energy is expected to dominate the relaxation to the bulk by considering the resemblance of the $\pi^{*}$ band. In the $\pi$ band, the carrier relaxation time to the bulk states should be shorter than that in the $\pi^{*}$ band because the  $\pi$ band is a resonance to the bulk states around $\overline{\Gamma}$. 

When we assume the following semi-classical model of the hot electron propagation in the dimer-axis direction, we can reproduce the observed linear relation between $L_c$ and $W$. The validity of the model means that the $\pi^{*}$ electron has a quantum coherence length shorter than a few nm in the dimer-axis direction. The phonon emission with the short interval may destroy the coherence during the slow propagation of the hot electron in this direction. 

For the analysis, we simply suppose that the hot electron loses energy to less than the threshold of the kink motion with a certain probability $q$ during the electron transfer from a dimer row to the adjacent dimer row. Furthermore, we presume that the $q$ for the $c$(4$\times$2) structure, $q_{c(4\times2)}$ is different from that for $p$(2$\times$2), $q_{p(2\times2)}$. 
Now, we consider the mixed surface structure which consists of $m_{c(4\times2)}$ dimer rows of the $c$(4$\times$2) structure and $m_{p(2\times2)}$ dimer rows of the $p$(2$\times$2) structure between the kink and the tunneling point as in Fig.4(a-d).  Then, the probability $P(m_{c(4\times2)},m_{p(2\times2)})$ that at least one electron moves the kink through ($m_{c(4\times2)}+m_{p(2\times2)}$) dimer rows after injecting $N$ electrons can be written as 

\begin{multline}
P(m_{c(4\times2)},m_{p(2\times2)}) \\
= 1 - (1 - s(1 - q_{c(4\times2)})^{m_{c(4\times2)}}(1 - q_{p(2\times2)})^{m_{p(2\times2)}})^N
\label{eq}
\end{multline}

\noindent
Here, $s$ is the probability that a single hot electron at the kink moves it. We used the assumption that each motion of the hot electron between the dimer rows is independent in the semi-classical model. The dotted curves in Figs. 3(a,b) are those fitted to this equation using the following parameters; s = $8.7\times10^{-9}$, $q_{c(4\times2)}$ = 0.42 and $m_{p(2\times2)}$ = 0  for  (a) and $s$ = $2.2\times10^{-8}$, $m_{c(4\times2)}$ = 0 and $q_{p(2\times2)}$ = 0.21 for (b) with the fixed $N$ = $1\times10^{11}$ for the both. It is noted that the parameter s should depend on the type of the kink.

To show the linear relation between $L_c$ and $W$, we define $m_{c} \equiv m_{c(4\times2)}$ and $m_{p} \equiv m_{p(2\times2)}$ at $L_c$, that is, $P(m_{c},m_{p}) = 0.5$, $L_c =  a(m_{c} + m_{p})$, $W = am_{c}$, and $a$ is the period of the dimer row. Then, the relation, $ m_{c}/m_{c0} + m_{p}/m_{p0}$ = 1, is obtained after calculations using eq. 1.
Here, $m_{c0}$ and $m_{p0}$ are experimentally determined on the pure $c$(4$\times$2) and $p$(2$\times$2) surfaces by the equations, $P(m_{c0},0) = 0.5$ and $P(0,m_{p0}) = 0.5$. The linear relation can be written as $L_c = - (m_{p0} - m_{c0})W/m_{c0} + am_{p0}$. Using the observed values $am_{c0}$ = 10.4 nm, and $am_{p0}$ = 27 nm in Figs. 3(a,b), we obtain, $L_c = - 1.6W + 27 $ nm. The coefficient of the linear term agrees with the slope of the dotted line in Fig. 4(e) while the constant term is higher than the experimental value. The discrepancy is attributed to the simple assumption in the semi-classical model. When the width of the $c$(4$\times$2) area is small, a quantum coherence of electrons may not be neglected during the propagation in the dimer-axis direction. 

The reason for the preferred direction of the kink motion is not clear.  The results shown in Fig. 2 indicate that the direction depends on the distance from the tunneling point and the bias voltage. The interaction between the electric dipole of the Ge dimer and the electric field due to $V_b$ is a possible origin of the preference because it depends on the distance from the tunneling position. The kink motion is essentially the inversion of the buckling orientation of the Ge dimer, and the electric dipole energy in the electric field can exceed the energy difference per dimer between $p$(2$\times$2) and $c$(4$\times$2).  For Si(001) surface, the $p$(2$\times$2) structure observed for positive $V_b$ is actually attributed to a similar electrostatic effect on the basis of density functional theory (DFT) \cite{Seino}. However, the observed different dependence on $V_b$ between the two types of the kink in Figs. 1(a,b) cannot be explained using only electric field. Moreover, the calculations by DFT under electric field are controversial \cite{Nakamura}. Another possible origin is the atomic force from the STM tip. A long-range attractive force between the tip and the surface \cite{Hoelscher} causes a local lattice deformation of the Ge substrate over a few tens nm in diameter, and can modify the potential barrier of the dimer flipping. 

In conclusion, we demonstrate that the hot carriers injected from the STM tip to the surface states induce systematic motions of surface dimer atoms apart from the tunneling point. In the case of the electron injection, the critical distance of the kink motion is anisotropic, and exceeds 80 nm in the dimer-row direction. It also depends on whether the surface superstructure is $p$(2$\times$2) or $c$(4$\times$2). These measurements illustrate the dynamics of the injected hot carriers in the surface electronic states. The observed anisotropy is consistent with the surface electronic structure. The semi-classical model of the hot electron propagation in the dimer-axis direction can reproduce the observed $L_c$ which linearly depends on the width of the $c$(4$\times$2) area between the B-type kink and the tunneling point.


\begin{thebibliography}{99} 
\bibitem{Eigler} D. M. Eigler, C.P. Lutz and W.E. Rudge: Nature (London) \textbf{352} (1991) 600.
\bibitem{Lyding} J. W. Lyding, T. C. Shen, J. S. Hubacek, J. R. Tucker and G. C. Abelin: Appl. Phys. Lett. \textbf{64} (1994) 2010.
\bibitem{Stipe} B. C. Stipe, M. A. Rezaei, W. Ho, S. Gao, M. Persson, and B. I. Lundqvist: Phys. Rev. Lett. \textbf{78} (1997) 4410. 
\bibitem{Maeda} Y. Nakamura, Y. Mera, and K. Maeda: Phys. Rev. Lett. \textbf{89} (2002) 266805.  
\bibitem{Weaver} K. S. Nakayama, E. Graugnard and J. H. Weaver: Phys. Rev. Lett. \textbf{89} (2002) 266106.  
\bibitem{Takagi4} Y. Takagi, Y. Yoshimoto, K. Nakatsuji and F. Komori: Surf. Sci. \textbf{559} (2004) 1.
\bibitem{Naitoh} Y. Naitoh, K. Nakatsuji and F. Komori: J. Chem. Phys. \textbf{117} (2002) 2832.
\bibitem{Takagi1} Y. Takagi, Y. Yoshimoto, K. Nakatsuji and F. Komori: J. Phys. Soc. Jpn. \textbf{72} (2003) 2425.
\bibitem{Kawai1} H. Kawai, Y. Yoshimoto, H. Shima, Y. Nakamura and M. Tsukada: J. Phys. Soc. Jpn. \textbf{71} (2002) 2192.
\bibitem{Kawai2} H. Kawai and O. Narikiyo: J. Phys. Soc. Jpn. \textbf{73} (2004) 2362.
\bibitem{Landermark1} E. Landemark, R.I.G. Uhrberg, P. Kruger and J. Pollmann: Surf. Sci. \textbf{236} (1990) L359.
\bibitem{Landermark2}E. Landemark,C. J. Karlsson, L.S.O Johansson and R.I.G. Uhrberg: Phys. Rev. B\textbf{49} (1994) 16523.
\bibitem{Kipp} L. Kipp, R. Manzke and M. Skibowski: Solid State Commun. \textbf{93} (1995) 603.
\bibitem{Nakatsuji} K. Nakatsuji, Y. Takagi, H. Kusuhara, A. Ishii and F. Komori: submitted.
\bibitem{Rohlfing} M. Rohlfing, P. Kr\"uger, and J. Pollmann: Phys. Rev. B\textbf{54} (1989) 13759.
\bibitem{Weinelt} M. Weinelt, M. Kutschera, R. Schmidt, C. Orth, T. Fauster and M. Rohlfing: Appl. Phys. A\textbf{80} (2005) 995.
\bibitem{Voelkmann} C. Voelkmann, M. Reichelt, T. Meier, S.W. Koch and U. Hofer: Phys. Rev. Lett. \textbf{92} (2004) 127405.
\bibitem{Seino} K. Seino, W.G. Schmidt and F. Bechstedt: Phys. Rev. Lett. \textbf{93} (2004) 036101.
\bibitem{Nakamura} J. Nakamura and A. Natori: Phys. Rev. B\textbf{71} (2005) 113303.
\bibitem{Hoelscher}H. H\"olscher, U. D. Schwarz, and R. Wiesendanger: Appl. Surf. Sci. \textbf{140} (1999) 344.

\end{thebibliography}
\end{document}